\documentclass[a4paper,12pt,epsfig]{article}
\usepackage{latexsym}
\usepackage{graphicx}
\begin{document}

\begin{center}

{\large\bf MeV ion-induced strain at nanoisland-semiconductor surface and interfaces}

\vspace{0.3in}
J. Ghatak, B. Satpati, M. Umananda, P. V. Satyam\footnote{To whom correspondence
 should be addressed: Tel: +91-674-2301058, Fax: +91-674-2300142; E-mail: satyam@iopb.res.in}

\vspace{0.1cm}
Institute of Physics, Sachivalaya Marg, Bhubaneswar, 751005, India

\vspace{0.5cm}
K. Akimoto, K. Ito

\vspace{0.1cm}
Department of Quantum Engineering, Nagoya University, Nagoya 464-8603, Japan

\vspace{0.5cm}
T. Emoto

\vspace{0.1cm}
Toyota National College of Technology, 2-1, Toyota, Aichi 471-8525, Japan
\end{center}
\vspace{1cm}

\begin{abstract}

Strain at surfaces and interfaces play an important role in the optical and 
electronic properties of materials. MeV ion-induced strain determination in 
single crystal silicon substrates and in Ag (nanoisland)/Si(111) at surface 
and interfaces has been carried out using transmission electron microscopy 
(TEM) and surface-sensitive X-ray diffraction. Ag nanoislands are grown under
various surface treatments using thermal evaporation in high vacuum conditions. Irradiation has been carried out with 1.5 MeV Au$^{\mathrm{2+}}$ ions at 
various fluences and impact angles.  Selected area electron diffraction 
(SAED) and lattice imaging (using TEM) has been used to determine the strain
at surface and interfaces. Preliminary results on the use of surface-sensitive asymmetric x-ray Bragg reflection method have been discussed. The TEM results 
directly indicate a contraction in the silicon lattice due to ion-induced 
effects. The nanoislands have shadowed the ion beam resulting in lesser strain
beneath the island structures in silicon substrates. High-resolution lattice
imaging has also been used to determine the strain in around amorphization 
zones caused by the ion irradiation.

\end{abstract}

\vspace{0.5cm}


\newpage

\section{Introduction}

Lattice strain at surface and interfaces plays an important role in the 
structural evolution of many material systems. Lattice strain may be caused 
by lattice defects or heteroepitaxy or other kind of stresses.  Knowing the 
strain distribution would result in tailoring the optical and electrical 
properties like the materials used in band gap engineering. In III-V compound
semiconductors the strain is very sensitive to the composition of two 
semiconductor layers and band gap can be tailored by varying the strain
between the layers \cite{1}. The coherent island growth in the epitaxial 
film growth (known as S-K growth mode) is known to be due to the strain 
relaxation in the growing film after an initial wetting layer \cite{2}. Hence, 
the degree of strain relaxation could be one of the main governing factors 
in coherent island growth. Strain measurement for individual islands would 
help to understand the coherent island growth mechanisms \cite{3}. In 
electronic device technology, the presence of localized stress fields at the
perimeter of the components of an integrated circuit found to play a negative 
role in obtaining the required device characteristics \cite{4}. A. Armigliato
et al., reported strain determination in silicon-based submicrometric electron
devices using electron microscopy methods \cite{5}. Recently, it was shown 
about the possibility of replicating nanostructures on silicon by low energy
ion beams by etching amorphized zones that were not shadowed by nanostructures
during ion beam irradiation \cite{6}.

In general, strain measurements are carried out by X-ray diffraction (XRD), 
transmission electron diffraction (TED) and Rutherford backscattering 
spectrometry/channeling (RBS/C) methods. These experimental techniques measure
the strain values averaged over length scales from sub-microns to few hundred
of microns. It is possible to get from the smaller length scales (in nanometer
scales) by using special geometries and conditions in the above experimental 
methods. For example, it is possible to limit the penetration of x-rays by
using grazing incidence methods or using RBS/C in glancing angle geometry. 
It is also possible to get strain in specific locations using nanobeam 
diffraction (NBD in TEM) or convergent beam diffraction (CBED in TEM) or 
lattice imaging (HRTEM).  

Among the X-ray methods, high resolution XRD or rocking curves is a common 
method to determine the strain distribution in single crystalline samples. 
The strain component is derived by simulating the rocking curves with a
model distribution of strains and by using dynamical theory formalism based
on Takagi-Toupin formalism \cite{7}. The rocking curve measurements have been 
carried out in grazing incidence geometry so as to make the method surface-
sensitive. Using this technique, known as asymmetric Bragg-rocking curves, 
one can quantitatively determine the strain by measuring the bulk reflection
under the grazing conditions \cite{8, 9, 10, 11}. T. Emoto et al., determined the strain due to nickel diffusion into hydrogen-terminated Si(111) surface using 
this method \cite{10}. Kishino et al., \cite{11} extended the dynamical 
diffraction theory derived by Battermann and Cole \cite{12} to the case of 
grazing incidence. Hasegawa et al., found the width of the rocking curve for 
(311) reflection depends on the width of the oxide layer thickness and 
reported strain field distribution at SiO$_{\mathrm{2}}$/Si(001) interface 
using the asymmetric Bragg reflection method \cite{13}. The x-ray methods 
are useful when the specimen under study remains mostly crystalline. The 
aforementioned X-ray methods give the information on integrated signal 
through the depth along which X-rays have penetrated. Similarly, Rutherford 
backscattering spectrometry/Channeling (RBS/C) is method to measure tetragonal
strain in the single crystalline systems. The RBS/C methods has been used to 
determine the tetragonal strain at CoSi$_{\mathrm{2}}$/Si(111) interface by
using axial channeling conditions. Previously, by combining HRXRD (which was
used to determine the strain component perpendicular to the surface normal)
and RBS/C (which was used to determine the tetragonal strain), the in-plane
strain component at CoSi$_{\mathrm{2}}$/Si(111) interface had been obtained 
\cite{14}. RBS/C experimental method also represents an integrated strain 
component.

Using TEM, both imaging mode (i.e., image contrast and lattice imaging) and 
diffraction mode (selected area diffraction (SAD), nano-beam diffraction (NBD) 
and large angle convergent beam diffraction (CBED)) have been exploited for 
the strain determination \cite{3, 5, 15, 16, 17, 18, 19, 20}. The strain was 
measured by comparing the image contrast predicted by the numerically 
integrated two-beam Howie-Whelan equations with the two-beam diffraction 
contrast images of strained copper matrix with coherent cobalt inclusions 
\cite{15}. The strain measurements on quantum dots have been relied upon 
strain simulations using finite element analysis (FE) due to lack of 
existence analytical strain models \cite{16}. Recently, a quantitative 
characterization of size, shape and the strain of a coherent island in 
semiconductor heterostructures have been determined by using bright-field 
suppressed diffraction imaging condition for planar-section specimen 
\cite{3,17}. Strains in crystals with amorphous surface films have been
studied by using CBED and HRTEM experimental methods \cite{18}.  Y. 
Androussi et al., reported the usefulness of fringe spacing in Moire-like
fringes in TEM images to determine the strain field at the apex of 
coherently strained islands and mean composition of the island \cite{19}. A. 
Armigliato et al. used the CBED technique, for strain determination in 
silicon-based submicrometric electron devices \cite{5}.

For last few decades in the fabrication process of integrated circuit devices,
ion implantation doping has become a primary step. The initial technological
driving force for MeV ion implantation was to form a deep conducting layer in
silicon. There had been several studies reporting on the MeV ion-beam-induced
damage studies in single crystal silicon \cite{21, 22, 23}. Energetic ions 
induce damage in the silicon at higher fluences, a phase transformation from 
crystalline Si (c-Si) to amorphous Si (a-Si) can occur. Damage induced by ion 
irradiation in Si depends on fluence, flux, energy, mass of the ion, target 
temperature, impact angle etc. Understanding the amorphization process is 
still an active area of research and various mechanisms have been put forward
\cite{24}. J. Kamila et al., reported that onset of amorphization due to MeV Au ion implantation in silicon occurs at lower fluence for low incident ion currents \cite{23}.  It could be interesting to have a systematic study to determine 
the strain evolution in the progression of amorphization. Previous strain 
measurements for the cases of 160 keV O$^{\mathrm{+}}$ ion and 1 MeV 
O$^{\mathrm{+}}$ ion irradiation in silicon show negative strain \cite{25,26}. 
In this article, determination strain profile will be presented using HRTEM
and asymmetric Bragg reflection methods.

Recently, a method of replicating nanostructures on Si surfaces using metal 
nanoparticles on Si as mask and low energy ion-irradiation has been proposed 
\cite{6}. Accordingly, Si nanostructures produced on Si substrate have a 
one-to-one correspondence with the self-assembled metal (Ag, Au, Pt) 
nanoparticles initially grown on the substrate. The smallest structures of 
Si thus produced emit red light when exposed to UV light \cite{6}. It is very important to understand the dominating factors for the amorphization process underneath the nanoislands. In this paper, shadowing effect of Ag nanoisland on the
substrate silicon will be presented.

\section{ Experimental}

The mirror polished (111) oriented Si single crystals were cleaned with 
de-ionized water followed by rinsing in methanol, trichloroethelene, 
methanol and a final rinse in de-ionized water. About 2.0 nm thick native 
oxide was present on the silicon surface. Implantations were carried out with 1.5 MeV Au$^{\mathrm{2+}}$ ions using 3.0 MV pelletron accelerator facility at 
Institute of Physics, Bhubaneswar.  The implantation were carried out room 
temperature with an incident ion current $\approx$ 20 nA.  To understand the 
shadowing effects of nanoparticles, isolated silver nanoislands were grown by
depositing 2 nm thick Ag film on Si(111) surface using resistive heating method in high vacuum conditions and later, the ion irradiations were carried out at 
room temperature at various fluences and impact angles. The fluences on the 
samples were varied from $1\times 10^{\mathrm{12}}$ to $1\times 10^{\mathrm{14}}$ ions cm$^{\mathrm{-2}}$. The substrates were oriented 5$^{\mathrm{o }}$ off 
normal to the incident beam to suppress the channeling effect for normal 
implantations (0$^{\mathrm{o}}$). Cross-sectional Transmission Electron 
Microscopy (XTEM) observation was carried out using a 200 kV JEOL JEM 2010 
(ultra high resolution mode) microscope for high resolution imaging. XTEM 
samples were prepared by a combination of mechanical polishing and followed 
by ion milling with 3.5 keV Ar$^{\mathrm{+}}$ ions. ImageJ \cite{27} is used
to analyze the data. Surface sensitive asymmetric X-ray diffraction 
experiments were carried out in atmosphere at room temperature at beam line
BL -15C at the Photon Factory of High Energy Accelerator Research Organization
in Tsukuba, Japan \cite{9}. For the observation of strain field near the 
substrate surface, the rocking curves of the (113) reflection of the substrate were measured. Since \{113\} planes are oriented at $\approx 29.5^o$ to \{111\} 
surface, the detector (NaI) was positioned at $\approx 59^o$ with respect to 
surface.  The incident X-ray energy was tuned such that the measurements were 
done near the critical angle of total reflection (i.e., grazing incidence 
geometry).

\section{Results and discussions}

TEM measurements have been used to determine the strain quantitatively. 
Preliminary data using x-ray asymmetric Bragg-reflection has been presented. 

Two TEM modes: (i) selected area diffraction and (ii) lattice imaging have
been used to determine the strain value from cross-section TEM specimen. The
selected area electron diffraction (SAED) has been used with lowest aperture
size (corresponding to $\approx$ 130 nm at the specimen) at different depths 
before and after irradiation to determine the strain distribution along the 
depth of the target.

 In general, the strain (in \%) is defined as: 
\begin{equation}
\epsilon=\frac{d-d_0}{d_0}\times 100
\end{equation}
where $\epsilon$ is the strain in percentage, d and d$_{\mathrm{0}}$ are the inter-planar spacing for strained and virgin specimen, Using  \textit{Dd =
L$\lambda$ } (for diffraction in TEM) , which can be written as
\begin{equation}
\epsilon=\frac{D_0-D}{D}\times 100
\end{equation}

where, D is the distance between the direct beam and diffracted beam for 
implanted sample (strained lattice) and D$_{\mathrm{0}}$ is that of the 
virgin sample, L is the camera length and $\lambda$ is the wavelength of  
electron.

Figure 1(a) shows the XTEM bright field image of silicon after irradiation
with 1.5 MeV Au$^{\mathrm{2+}}$ at a fluence $5\times 10^{\mathrm{13}}$ ions 
cm$^{\mathrm{-2}}$. Diffraction patterns from three selected areas 
corresponding to various depths have been taken: surface, end of range (EOR) 
and bulk (un-irradiated part). The selected area at the specimen (SAED aperture 
corresponds to a region $\approx$ 130nm) at various depths corresponding to the above areas is show as circles with I, II and III legends inside the circles in 
the BF image.  The respective diffraction pattern from surface (I), EOR (II) 
and bulk (III) have been shown in Figures 1(b), (c) and (d) respectively. From
the SAED pattern, it is clear that a complete amorphization has not taken 
place even at EOR. It was shown in our previous work, the fluence required for
amorphization was $1\times 10^{\mathrm{14}}$ ions cm$^{\mathrm{-2}}$ \cite{23}. Care had been taken to avoid the overlap of regions and alignment of aperture 
while taking SAED. The values of D and D$_{\mathrm{0}}$ of both implanted and un- implanted (not shown) silicon have been determined. Strain in percent at 
different regions for different set of planes has been presented in Table 1.
From the data of Table 1, it is clear that after the irradiation negative 
strain i.e., lattice contraction is induced in the system, similar to the 
work by S. L. Ellingboe et. al.,\cite{26} at very high fluence compared to our 
work. At the EOR the strain is maximum for all set of planes and strain at 
surface is relatively weaker than other irradiated places. This is due to 
the damage and defect density at the EOR.

\begin{table}[htbp]
\caption{Strain (\%) at different depths of different set of planes for
silicon irradiated with 1.5 MeV Au$^{2+}$ at a fluence of $5\times 10^{13}$ ions cm$^{-2}$ (Typical error in $\epsilon$ is $<$ 1\%)}
\begin{center}
\begin{tabular}{|c|c|c|c|}
\hline
Set of planes& Strain at surface & Strain at EOR&Strain at Bulk \\
\hline
\{002\}&-0.74 &-1.47&-0.74  \\
\hline
 \{-111\}&-0.86  &-1.70&-1.28 \\
\hline
\{-11-1\}& -0.81&-1.20 &-0.81 \\
\hline
\{-220\}&0.49&-0.75&-0.25\\
\hline
\end{tabular}
\end{center}
\end{table}

Figure 2(a) and (b) show the XTEM high-resolution bright field image at 
surface and EOR for post-irradiated Si with 1.5 MeV Au$^{\mathrm{2+}}$ at a 
fluence of $5\times 10^{\mathrm{13}}$ ions cm$^{\mathrm{-2}}$. It is clear 
from the figure that the surface after implantation remains crystalline. 
Previous results showed the amorphization to occur at $1\times 10^
{\mathrm{14}}$ ions cm$^{\mathrm{-2 }}$ when the irradiation was carried 
out at an impact angle of 60$^o$ \cite{23}. From the Figure 2(b), amorphous 
zones are present at EOR region. The interface of amorphous/ crystalline 
(a/c) is found to be sharp \cite{28}. In this region, the d-spacing at
various depths of the substrate has been measured from the lattice 
images. From the diffraction data, the strain is found to be maximum
at EOR. We have found that the d spacing away from amorphous regions 
is same as that at the surface and differs compared to the regions 
around amorphous zone. With the increase of fluence, the strain at the 
a/c interface found to increase and hence could result in amorphization at
sufficient  high fluence.

The figure 3(a) shows the XTEM bright field image of as deposited Ag 
nanoislands on Si(111) substrate. Figure 3(b) and 3(c) show the XTEM bright
field image corresponds to with and without Ag nanoislands after the ion 
irradiation at a fluence of $1\times 10^{\mathrm{13}}$ ions cm$^{\mathrm{-2}}$. 
In this case also, the strain has induced to the system after ion irradiation.
Without Ag islands, the strain in Si surface is similar to that of Figure 2(a). But with the Ag islands, the strain in Si surface is found to be negligible
and found to retain its crystalline structure even at larger fluences. That
means the islands shadow the effect of irradiation on the surface. However the
projected range of Au ions in Ag ($\approx$ 130 nm calculated from TRIM 
\cite{30}) much greater than the average height of the Ag islands ($\approx 
$17 nm). So, in the same sample with Ag islands, amorphization would occur at
higher fluence at different depths of the Si sample. The places without 
islands are getting amorphized at lower fluence due to lower strain
values.

The experimental rocking curve obtained under asymmetric Bragg condition is
shown (Figure 4) for virgin Si(111), 1.5 MeV Au$^{\mathrm{2+}}$ ion implanted
at 5$^o$ impact angle  at a fluence of  $5\times 10^{\mathrm{12}}$ and at 
fluence $5\times 10^{\mathrm{13}}$ ions cm$^{\mathrm{-2}}$. From the Figure 4,
the rocking curve from the irradiated specimen at a  fluence of $5\times 10^
{\mathrm{13}}$ ions cm$^{\mathrm{-2}}$ is found to be asymmetric and show 
equally spaced satellite peaks. If it is assumed that these secondary peaks 
arises due to strain, then the maximum strain for the peak at extreme right 
show a maximum of 0.1\% lattice contraction in \{113\} inter-planar spacing.
The appearance of oscillations (or satellite peaks) arises either due to strain or due to the presence of a very thin layer of different electron density from the substrate matrix. Simulations using modified Darwin theory and dynamical 
diffraction theory indicate broadening of rocking curve and appearance of 
satellite peak due to presence of compressive strain in the system \cite{10}.
A detailed analysis is under progress to understand x-ray data on various set
of samples. 

\section{Conclusion}

We have shown that strain can be determined along the depth by electron 
diffraction (SAED) and high resolution lattice imaging (HRTEM). TEM 
measurements show a compressive strain due to MeV ion implantation. Strain 
is found to be maximum at end of range (EOR) and relatively weaker at the 
surface. With the increase of fluence, the volume of amorphous zone is 
increased and the substrate tends to become amorphize from crystalline 
structure. Ag nanoislands on the silicon substrate found to act as mask the 
underneath substrate resulting in retaining crystalline structure beneath the
islands. 

\section{Acknowledgements:}

We thank Mr. A.K. Dash for his help in TEM measurements and the staff at IBL,
Institute of Physics for ion implantations. We thank Dr. K. Hirano of the 
Photon Factory for help with X-ray measurements. P. V. Satyam would like to
thank JSPS, Japan for funding his visit to KEK, and Nagoya University, Nagoya,
Japan under invitation fellowship.

\newpage


{\large\bf Figure Captions}

\begin{figure}[htbp] 
\caption{\label{Fig1}(a) TEM bright field image of 1.5 
MeV Au$^{\mathrm{2+}}$ implanted Si(111) at a fluence of $5\times 10^{\mathrm{13}}$cm$^{\mathrm{-2 }}$and (b), (c) and (d) are the SAED pattern of Si(111) from
Surface, EOR and beyond EOR (bulk) respectively(corresponding regions I, II 
and III are depicted in the BF image).}
\end{figure}

\begin{figure}[htbp] 
\caption{\label{Fig2}HRTEM bright field image (a) at 
surface (b) at EOR of 1.5 MeV Au$^{\mathrm{2+}}$ implanted Si(111) with 
$5\times 10^{\mathrm{13}}$ cm$^{\mathrm{-2}}$.}
\end{figure}

\begin{figure} [htbp]
\caption{\label{Fig3}XTEM bright field image of  (a) as 
deposited Ag islands and (b) with and (c) without island irradiated with 1.5
MeV Au$^{\mathrm{2+}}$ at a fluence $1\times 10^{\mathrm{13 }}$ions 
cm$^{\mathrm{-2}}$ on same sample.  }
\end{figure}

\begin{figure}[htbp]
\caption{\label{Fig4}Experimental X-ray data for 
asymmetric diffraction from (113) planes of single crystalline Si (virgin, 
1.5 MeV Au ion irradiated at two fluences).}
\end{figure}


\begin{thebibliography}{}

\bibitem{1}S.T. Picraux, B.L. Doyle, J.Y. Tsao, Semiconductors and Semimetals, 
in: Thomas P. Pearsall (Ed.), Strained-Layer Superlattices: Materials Science 
and Technology, Academic Press, New York, 1991

\bibitem{2}	 I. Daruja, J, Tersoff, A. -L. Barabasi, Phys. Rev. Lett. \textbf{82} (1999) 330

\bibitem{3}	Chuan-Pu Liu, Murray Gibson, Thin Solid Films \textbf{424} 
(2003) 2

\bibitem{4}	S. M. Hu, J. Appl. Phys. \textbf{70 }(1991) R53

 
\bibitem{5}	A. Armigliato, R. Balboni, S. Balboni, S. Frabboni, A. Tixier,
G. P. Carnevale, P. Colpani, G. Pavia, A. Marmiroli, Micron \textbf{31} (2000)
203

\bibitem{6}	B. Satpati and B. N. Dev, Nanotechnology {\bf 6} (2005) 572.

\bibitem{7}	Chu Ryang Wie, Mater. Sci. and Engg. \textbf{R13 }(1994) 1

\bibitem{8}	T. Emoto, K. Akimoto, A. Ichimiya, J. Synchrotron Rad. \textbf{5} (1998) 964 
 
\bibitem{9}	T. Emoto, K. Akimoto. Y. Ishikawa, A. Ichimiya, A. Tanikawa, 
Thin Solid Films \textbf{369 }(2000) 281
 
\bibitem{10}	T. Emoto, K. Akimoto, A. Ichimiya, Surf. Sci. \textbf{438} 
(1999) 107  
 
\bibitem{11}	S. Kishino, K. Kohra, Jpn. J. Appl. Phys.\textbf{ 10} (1971) 
55
  
\bibitem{12}	B. W. Battermann and H. Cole, Rev. Mod. Phys. \textbf{36} (1964) 681
  
\bibitem{13} 	 E. Hasegawa, A. Ishitani, K. Akimoto, M. Tsukiji, N. Ohta, J.
Electrochem. Soc. \textbf{142 }(1995) 273
  
\bibitem{14} 	P. V. Satyam, B. Sundaravel, S. K. Ghose, B. Rout, K. Sekar, D. P. Mahapatra, B. N. Dev, Indian Journal of Physics, Part A, Volume \textbf{70A} (1996) 783
  
\bibitem{15}  M. F. Ashby, L. M. Brown, Philos. Mag. \textbf{8} (1963) 1083 
  
\bibitem{16} 	T. Benabbas, P. Fran ois, Y. Androussi, A. Lefevbre, J. Appl.
Phys. \textbf{80} (1996) 2763 
  
\bibitem{17} 	 Peter D. Miller, Chuan-Pu Liu, J. Murray Gibson, 
Ultramicroscopy \textbf{84} (2000) 225
  
\bibitem{18} 	 F. Banhart, Ultramicroscopy \textbf{56} (1994) 233 
  
\bibitem{19} 	Y. Androussi, T. Benabbas, A. Lefevbre, Ultramicroscopy 
\textbf{93} (2002) 161 
  
\bibitem{20}	C. Bocchi, F. Germini, E. Kh. Mukhamedzhanov, L. Nasi, 
V.Privitera, C. Spinella, Mater. Sci. and Engg. \textbf{B91-92} (2002) 457 
  
\bibitem{21}	N. W. Cheung, C. L. Liang, B. K. Liew, R. H. Mutikainen,
H. Wong, Nucl. Instr. and Metho. \textbf{B 37/38} (1989) 941
 
\bibitem{22}	J. S. Williams, et al., Nucl. Instr. and Metho. 
\textbf{B 80/81} (1993) 507
  
\bibitem{23}	J. Kamila, B. Satpati, D. K. Goswami, M. Rundhe, B. N. Dev,
P. V. Satyam, Nucl. Instr. and Meth. \textbf{B 207} (2003) 291
  
\bibitem{24}	T. Motoka, S. Harada, M. Ishimaru, Phys. Rev. Lett. 
\textbf{78} (1997) 2980
  
\bibitem{25}	D. Venables, K. S. Jones and F. Namavar, J. Appl. Phys. 
\textbf{60} (1992) 3147
  
\bibitem{26}	S. L. Ellingboe and M. C. Ridway, Nucl. Instrum. Meth.  
Phys. Res. \textbf{B 106} (1995)
  
\bibitem{27} http://rsb.info.nih.gov/ij/
  
\bibitem{28}	T. Motooka, S. Hirada and M. Ishimaru, Phys. Rev. Lett. 
\textbf{78} (1997) 2980
  
\bibitem{29}	J. Narayan, D. Fathy, O. S. Oen and O. W Holland, J. Vac. 
Sci. Technol. \textbf{A} \textbf{2} (1984) 1303
  
\bibitem{30}	J. B. Biersack, L. G. Haggmark, Nucl. Instrum. Methodhs 
Phys. Res.\textbf{174} (1980) 257
\end{thebibliography}
\end{document}